\documentclass[floatfix,pra,amsmath,amssymb,letterpaper,groupaddresses,superscriptaddress,twocolumn]{revtex4}
\usepackage{times}
\usepackage{latexsym}
\usepackage{graphicx}
\usepackage{amsmath,amssymb}
\usepackage{verbatim,times,bbm}
\usepackage{subfigure}
\usepackage{soul}
\usepackage[colorlinks,hyperindex]{hyperref}
\hypersetup{colorlinks,citecolor=blue,linkcolor=blue,urlcolor=blue}
\usepackage{float}
\usepackage{color}

\usepackage{booktabs,multirow,array,hyperref}

\begin{document}
	
\title{Object Detection Using Quantum Transfer Learning}

\author{Mohammad Bahrami\footnote{mohammadbahrami@shahroodut.ac.ir}}
\affiliation{Faculty of Physics, Shahrood University of Technology, 3619995161 Shahrood, Iran}

\author{Morteza Rafiee\footnote{m.rafiee@shahroodut.ac.ir}}
\affiliation{Faculty of Physics, Shahrood University of Technology, 3619995161 Shahrood, Iran}

\author{Mohsen Norouzi\footnote{m.norouzi@shahroodut.ac.ir}}
\affiliation{Faculty of Computer Engineering, Shahrood University of Technology, 3619995161 Shahrood, Iran}

\begin{abstract}
Object detection requires the joint learning of semantic identity and continuous spatial geometry. Existing quantum transfer learning approaches have focused mainly on classification tasks and do not support regression. Here, we introduce a physics informed hybrid classical-quantum architecture for object detection in which a pre-trained MobileNetV2 maps visual features into an eight qubits variationall quantum Hilbert space. The Hilbert space is partitioned into two task specific subspaces for semantic classification and geometric localization. Local quantum expectation values are used to predict continuous bounding box coordinates, while circuit topology regulates the flow of quantum information between the two tasks. We compare linear and circular CNOT topologies and show that periodic boundary conditions suppress bipartite entanglement without degrading detection performance. The circular architecture reaches a bipartite von Neumann entropy of 0.1688 bits and a Meyer-Wallach entanglement measure of 0.0370, while achieving an mAP@0.5 of 0.9448 and an mAP@0.5:0.95 of 0.6558. At the same time, the model uses only 10,366 trainable parameters, corresponding to a 99.31\% reduction relative to the large classical baseline, while maintaining nearly equivalent detection performance, including comparable performance under the more stringent mAP@0.5:0.95 criterion. A thermodynamic formulation further interprets the total learning loss as a variation in Helmholtz free energy. It interprets geometric localization as an effective internal energy contribution and semantic classification as an entropic contribution. These results indicate that high performance in the proposed object detection model does not require maximal entanglement. Instead, efficient learning can emerge through topology controlled information flow, targeted restriction of entanglement, and information renormalization within the Hilbert space.
\end{abstract}
\maketitle

\section{Introduction}
The convergence of quantum information physics and machine learning provides a new framework for addressing computationally demanding problems in artificial intelligence. Object detection is particularly challenging because it requires the joint learning of semantic identity and continuous spatial geometry. In classical multi-task learning, these objectives rely on different inductive biases. Semantic classification requires translation invariance, whereas geometric localization requires translation equivariance \cite{ref1}.

Classical neural networks may struggle to represent these objectives simultaneously in high-dimensional feature spaces. Their inductive biases may not be sufficient to disentangle the nonlinear correlations associated with semantic and geometric information \cite{ref2}. Increasing model capacity through overparameterization can partly address this limitation \cite{ref3}. However, this strategy substantially increases the number of trainable parameters and may lead to significant performance degradation when the model is aggressively compressed.

Quantum machine learning (QML) provides an alternative computational framework for the Noisy Intermediate Scale Quantum (NISQ) era. However, practical quantum models remain limited by trainability, circuit depth, and barren plateaus. In sufficiently deep circuits, gradient variance can decrease exponentially with the number of qubits. This effect becomes particularly relevant when the circuit approaches approximate unitary 2-designs \cite{ref4}.

Most quantum transfer learning (QTL) approaches have focused on classification tasks. They commonly treat the quantum circuit as a nonlinear processing module. The roles of Hilbert space geometry, circuit topology, and entanglement structure in organizing learned representations have received less attention. These limitations become more important in object detection, where semantic classification and continuous bounding box regression must be optimized jointly. Such joint optimization can introduce task interference and negative transfer.

Here, we introduce a physics informed hybrid classical-quantum framework that extends QTL to object detection. A pre-trained MobileNetV2 backbone extracts high-dimensional visual features. These features are then mapped to an 8 qubit variational quantum circuit through a trainable compression and angle encoding layer.

The Hilbert space is partitioned into two task specific subspaces. Qubits 0 to 3 encode semantic information for classification. Qubits 4 to 7 encode geometric information for continuous localization. Local Pauli Z expectation values serve as quantum observables for classification and prediction of the bounding box coordinates. This design extends QTL beyond conventional classification toward Hybrid Quantum Regression.

The circuit topology is used to regulate information flow and entanglement between the two task specific subsystems. Linear and circular CNOT topologies are compared to examine the effects of open and periodic boundary conditions. The proposed framework also incorporates observable learning. Trainable classical coefficients combine local quantum measurements to generate task specific readouts.

We further formulate the learning dynamics within a thermodynamic framework. The learning process can be interpreted through changes in Helmholtz free energy. Within this formulation, the GIoU loss contributes an effective internal energy term, whereas the cross entropy loss contributes an effective entropic term.

The classical compression layer can be interpreted as a discrete step in information renormalization. It reduces the dimensionality of the classical representation before quantum evolution. This formulation provides a unified description of semantic classification and geometric localization within the Hilbert space. It also connects the optimization trajectory to information flow and thermodynamic equilibration.

Our results show that high detection performance does not require maximal entanglement. The circular topology imposes periodic boundary conditions and reduces the bipartite von Neumann entropy from approximately 0.7 bits in the linear architecture to 0.1688 bits. The Meyer-Wallach entanglement measure also decreases to 0.0370.

The proposed circular architecture achieves an mAP@0.5 of 0.9448 and an mAP@0.5:0.95 of 0.6558. The model uses only 10,366 trainable parameters. This corresponds to a $99.31 \%$ reduction relative to the large classical baseline. Despite using the same backbone, the proposed model maintains nearly equivalent detection performance despite the substantial reduction in the number of trainable parameters in the prediction head.

These results indicate that the computational role of entanglement is not determined by its magnitude alone. Instead, efficient learning can emerge from topology controlled information flow, targeted entanglement restriction, and structured information renormalization within the Hilbert space.

The remainder of this paper is organized as follows. Section \ref{sec2} reviews quantum transfer learning and existing quantum approaches to object detection. Section \ref{sec3} presents the proposed hybrid architecture, the variational quantum circuit, the task specific partitioning of the Hilbert space, the quantum information metrics, and the thermodynamic formulation. Section \ref{sec4} evaluates detection performance and examines the information and thermodynamic dynamics of the model. Section \ref{sec5} concludes the study and discusses directions for future research.

\section{Related Works} \label{sec2}
The concept of quantum transfer learning (QTL) was introduced by Mari et al. \cite{ref5}. Their study demonstrated that features extracted by classical deep neural networks can be transferred to quantum circuits for subsequent processing. Later studies showed that parameterized quantum circuits (PQCs) can be integrated into the final layers of classical deep learning architectures. ResNet and MobileNet have been used for this purpose \cite{ref6}.

Most existing QTL approaches provide limited analysis of the underlying Hilbert space and its role in feature representation. In classical multi-task learning, semantic classification and geometric localization represent distinct learning objectives. Optimizing these objectives jointly can introduce task interference and may result in negative transfer  \cite{ref7}.

Table~\ref{tab:modalities} categorizes existing QTL studies across three data modalities: image, text, and audio. These studies cover a broad range of applications, including natural language processing and medical imaging. However, most existing QTL research has focused on classification tasks. More complex tasks, such as object detection, have received limited attention. Object detection requires both semantic classification and continuous bounding box regression. This combination introduces additional challenges for QTL frameworks, particularly in learning continuous geometric parameters and managing task interference within variational quantum circuits.

In this context, we extend the QTL framework beyond conventional classification tasks. The proposed framework integrates semantic classification and geometric localization within a unified architecture. This design extends QTL to object detection. It also provides a framework for jointly learning semantic and geometric information within a variational quantum model. Consequently, the proposed approach addresses an important gap in the current quantum machine learning literature.
 
\begin{widetext}
\begin{center}
\begin{table}[H]
	\centering
	\caption{Categorization of the reviewed works on machine vision using the quantum transfer learning model.}
	\label{tab:modalities}
	\renewcommand{\arraystretch}{1.10}
	\setlength{\tabcolsep}{4pt}
	
	\begin{tabular}{|p{2.2cm}|p{3.8cm}|p{5.5cm}|p{2.0cm}|}
		\hline
		\textbf{Data Modality} &
		\textbf{Application Domain} &
		\textbf{Specific Task(s)} &
		\textbf{References} \\
		\hline
		
		
		\multirow{9}{*}{Image}
		&
		\multirow{3}{*}{%
			\parbox{3.6cm}{\centering
				Computer Vision \&\\
				Pattern Recognition}}
		&
		Handwriting and digit classification
		&
		\cite{ref8,ref9,ref10}
		\\
		
		\cline{3-4}
		
		& &
		Anomaly detection, Image synthesis
		&
		\cite{ref11,ref12}
		\\
		
		\cline{3-4}
		
		& &
		Deepfake detection
		&
		\cite{ref13,ref14}
		\\
		
		\cline{2-4}
		
		&
		Remote Sensing \& Transportation
		&
		Aerial/satellite imaging, vehicle classification
		&
		\cite{ref15,ref16,ref17}
		\\
		
		\cline{2-4}
		
		&
		Environment
		&
		Scene classification
		&
		\cite{ref18,ref19}
		\\
		
		\cline{2-4}
		
		&
		\multirow{4}{*}{%
			\parbox{3.6cm}{\centering
				Medical Imaging \&\\
				Diagnostics}}
		&
		Oncology (various cancer detections)
		&
		\cite{ref20}
		\\
		
		\cline{3-4}
		
		& &
		Infectious diseases (COVID-19 detection)
		&
		\cite{ref21,ref22,ref23,ref24,ref25,ref26}
		\\
		
		\cline{3-4}
		
		& &
		Neurodegenerative and cardiovascular diseases
		&
		\cite{ref27,ref28,ref29,ref30}
		\\
		
		\cline{3-4}
		
		& &
		Knee osteoarthritis
		&
		\cite{ref31}
		\\
		
		\hline
		
		
		\multirow{2}{*}{Text}
		&
		Natural Language Processing (NLP)
		&
		Acceptability judgments
		&
		\cite{ref32}
		\\
		
		\cline{2-4}
		
		&
		Cybersecurity \& Social Media
		&
		SMS spam detection, Adverse drug reaction detection
		&
		\cite{ref33,ref34}
		\\
		
		\hline
		
		
		Multimodal (Text \& Image)
		&
		Representation Learning
		&
		Image-text matching
		&
		\cite{ref35}
		\\
		
		\hline
		
		
		Audio/Voice
		&
		Environmental Monitoring
		&
		Environmental Sound Classification (ESC)
		&
		\cite{ref36}
		\\
		
		\hline
		
		
		\textit{Image (Proposed Model)}
		&
		\textit{Advanced Computer Vision}
		&
		\textit{Object Detection (Simultaneous Semantic Classification
			and Continuous Geometric Regression)}
		&
		\textit{[This Paper]}
		\\
		
		\hline
		
	\end{tabular}
\end{table}
\end{center}
\end{widetext}
Statistical physics provides a useful framework for analyzing the behavior of neural networks. Tishby and Zaslavsky \cite{ref37} introduced the information bottleneck principle to describe the dynamics of deep learning. Their framework suggests that deep neural networks gradually suppress irrelevant information while preserving task relevant variables. This process can be interpreted as a renormalization flow in which the representation becomes more compact while retaining the information needed for the learning task.

In the quantum domain, Amin et al. \cite{ref38} developed quantum Boltzmann machines and established a connection between probability distributions in machine learning and thermal fluctuations. Subsequent studies have investigated thermodynamic formulations of quantum machine learning models. However, these approaches have focused mainly on binary classification tasks \cite{ref39}. Here, we extend this perspective to object detection. We formulate the learning process in terms of Helmholtz free energy within the Hilbert space. The geometric loss, defined by the generalized intersection over union (GIoU), represents the internal energy contribution. The cross entropy loss represents the entropic contribution. This formulation provides a uniﬁed thermodynamic interpretation of semantic classiﬁcation and geometric localization within a hybrid classical-quantum architecture.

One of the main challenges in scaling quantum neural networks is the barren plateau phenomenon. In sufficiently deep quantum circuits, the variance of the gradients can decrease exponentially with the number of qubits. McClean et al. \cite{ref40} showed that quantum circuits with unconstrained entanglement can approach approximate unitary 2-designs. As a result, the gradients can become exponentially small, which makes parameter optimization increasingly difficult.

Cerezo et al. \cite{ref41} showed that local observables can mitigate the exponential suppression of gradients. However, the role of entanglement is still often discussed in terms of maximizing the available entanglement capacity. This perspective does not fully account for how circuit structure and topology influence the learned representation \cite{ref42}. In particular, limited attention has been given to the use of circuit topology to control entanglement. This includes topologies with open and periodic boundary conditions.

In this work, we investigate targeted entanglement restriction. We show that circuit topology can be used to regulate entanglement during training. This approach improves parameter efficiency while operating within only a small fraction of the theoretical von Neumann entropy bound.

\begin{table}
	\centering
	\caption{Summary of the reviewed works on quantum models for object detection.}
	\label{tab:qod}
	\renewcommand{\arraystretch}{1.25}
	\setlength{\tabcolsep}{10pt}
	\begin{tabular}{lccccc}
		\toprule \hline \hline
		\textbf{Reference}
		& \multicolumn{5}{c}{\textbf{Feature}} \\
		\hline \hline \cmidrule(lr){2-6}
		& \textbf{I} & \textbf{II} & \textbf{III} & \textbf{IV} & \textbf{V} \\
		\midrule
		
		\cite{ref43} & $\times$ & $\times$ & $\times$ & $\checkmark$ & $\checkmark$ \\
		\cite{ref44} & $\times$ & $\times$ & $\times$ & $\times$ & $\times$ \\
		\cite{ref45} & $\times$ & $\times$ & $\times$ & $\checkmark$ & $\checkmark$ \\
		\cite{ref46} & $\times$ & $\times$ & $\times$ & $\checkmark$ & $\checkmark$ \\
		\cite{ref47} & $\times$ & $\times$ & $\times$ & $\checkmark$ & $\checkmark$ \\
		\cite{ref48} & $\times$ & $\times$ & $\times$ & $\checkmark$ & $\checkmark$ \\
		\cite{ref49} & $\times$ & $\times$ & $\times$ & $\times$ & $\checkmark$ \\
		\cite{ref50} & $\times$ & $\times$ & $\times$ & $\times$ & $\checkmark$ \\
		
		\midrule
		
		\textbf{This paper}
		& $\checkmark$ & $\checkmark$ & $\checkmark$ & $\checkmark$ & $\checkmark$ \\
		\hline
		\bottomrule
	\end{tabular}
	
	\vspace{4pt}
	
	\begin{minipage}{0.95\linewidth}
		\footnotesize
		\textbf{I:} Hybrid Quantum Regression;
		\textbf{II:} Topological Entropy Suppression;
		\textbf{III:} Task-Aware Hilbert Space Partitioning;
		\textbf{IV:} Parameter-Efficient Hybrid Architecture;
		\textbf{V:} Validation on High-Dimensional Datasets.
	\end{minipage}
\end{table}

Few studies have investigated quantum approaches to object detection. The existing methods are summarized in Table~\ref{tab:qod}. Recent studies have applied quantum computing to real visual datasets with high-dimensional feature spaces (Feature V). However, from a physics based perspective of machine learning, most of these approaches still follow the classical computational paradigm. Methods based on adiabatic quantum computing \cite{ref48,ref49,ref50} mainly address discrete optimization problems, such as graph assignment. These methods do not provide a general framework for learning continuous feature representations. Hybrid approaches, including \cite{ref43,ref45,ref46,ref47} typically use the quantum circuit as a feature extractor. Geometric localization is then performed by classical fully connected layers. As a result, these approaches do not support Hybrid Quantum Regression (Feature I). They may also be sensitive to destructive interference and negative transfer when semantic classification and geometric localization are optimized jointly.

The proposed architecture follows a different approach. It introduces a physics informed design for the NISQ era. The quantum circuit is not treated as a direct replacement for a classical module. Instead, semantic classification and geometric localization are modeled through the dynamics of the quantum state. Periodic boundary conditions are used to regulate topological entanglement (Feature II). The Hilbert space is partitioned according to the roles of the two tasks (Feature III). This structure enables hybrid prediction of bounding box coordinates from quantum expectation values (Feature I). It also improves parameter efficiency (Feature IV). These results indicate that high performance in a multi task quantum model does not require maximal entanglement. Instead, it can emerge from controlled information flow and targeted feature renormalization within the Hilbert space.

\begin{figure}[H]\centering\includegraphics[width=1\linewidth]{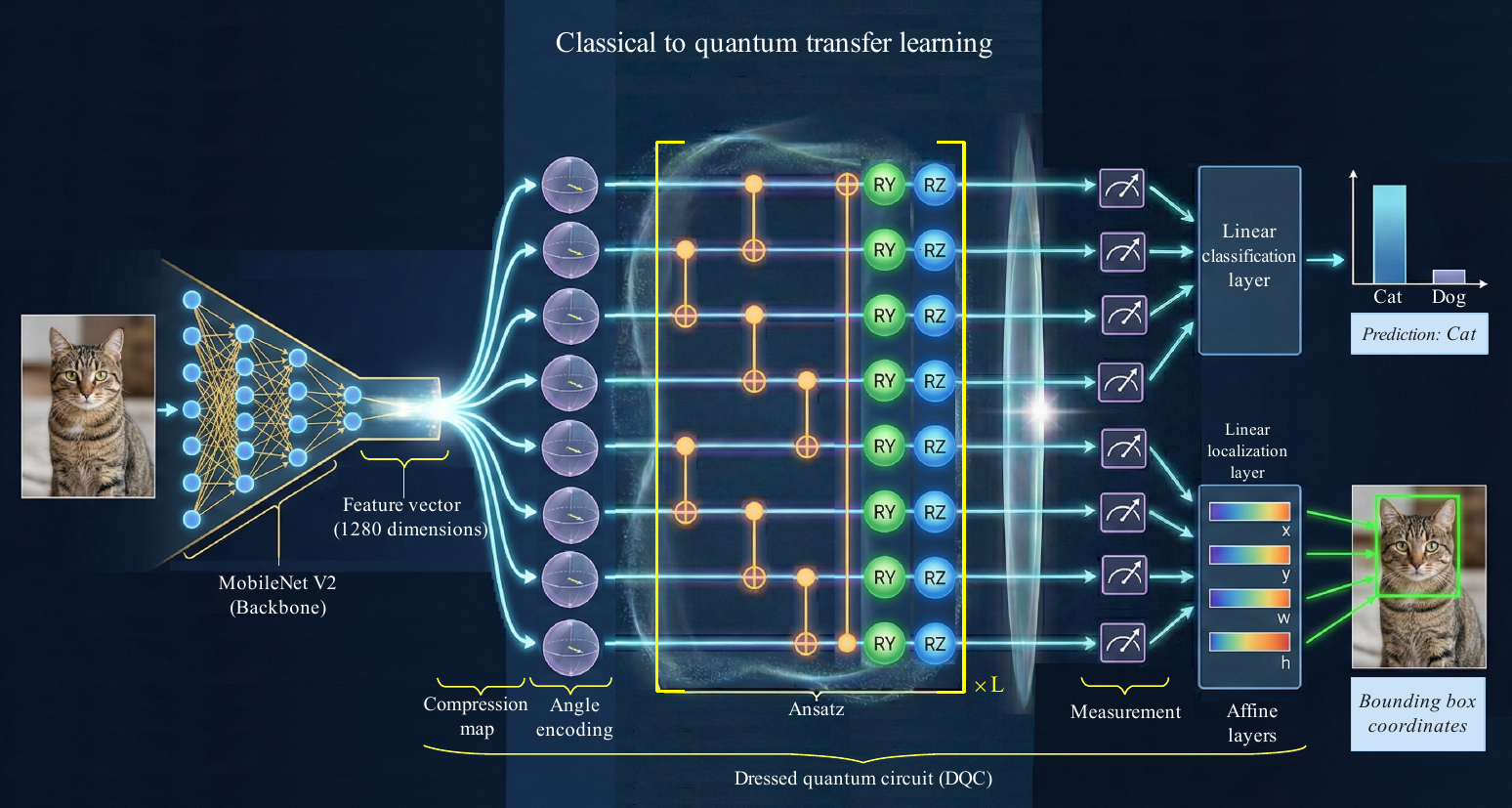}\caption{Architecture of the entanglement-enhanced hybrid classical-quantum object detection framework.}\label{fig:architecture}\end{figure}

\section{Proposed Method} \label{sec3}
The proposed architecture is a hybrid framework based on classical to quantum transfer learning. A pre-trained classical backbone is connected to a Dressed Quantum Circuit (DQC). As shown in Figure 1, the input images from the Kaggle dataset \cite{ref51} are first processed by a MobileNetV2 backbone. 
The pre-trained MobileNetV2 backbone is kept fully frozen throughout training. Visual features are extracted from the final convolutional representation after Global Average Pooling. This operation produces a feature vector with 1280 dimensions for each image. All Batch Normalization layers in the classical backbone are also kept permanently in evaluation mode during training. This prevents the normalization statistics from changing during optimization and helps maintain a stable feature distribution. These features are then mapped to the variational quantum circuit through a compression mapping and angle encoding.
 We formulate this process as a mapping from a high-dimensional classical feature manifold $\mathcal{M}\subset\mathbb{R}^{D}$ to a quantum Hilbert space $\mathcal{H}$. The entangling gates generate correlations among the qubits during the quantum evolution. In Figure~\ref{fig:architecture}, this entanglement is illustrated as a halo surrounding the qubits. The quantum state is then subjected to local measurements. The output space is partitioned into two subspaces associated with the two tasks. The first subspace is used for semantic classification, while the second is used for geometric localization. Following optimization, the model produces the object detection outputs for the input images.

\subsection{Classical Information Encoding}
In this work, we formulate object detection as an embedding problem. The classical feature distributions are mapped into a high-dimensional quantum Hilbert space, $\mathcal{H} \cong (\mathbb{C}^{2})^{\otimes N}$, where $N=8$ is the number of qubits. A pre-trained MobileNetV2 network is used as the classical backbone. We use MobileNetV2 not only because of its compact architecture, but also because it provides an efficient feature representation. The backbone extracts a classical feature vector $\mathbf{x}\in\mathbb{R}^{D}$, where $D=1280$.

The classical feature vector is first compressed before being encoded into the quantum circuit. A system with 8 qubits spans a Hilbert space with $2^{8} = 256$ computational basis states. The embedding process therefore maps the high-dimensional classical representation into a compact quantum representation. The resulting quantum state encodes information through the amplitudes and phases of the superposition.

We define the embedding layer as a compression map $\mathcal{R}:\mathbb{R}^{D} \longrightarrow \mathbb{R}^{N}$. This mapping connects the reduced classical representation to the variational quantum circuit. It projects the feature vector $\mathbf{x}\in\mathbb{R}^{1280}$ onto a compressed latent vector $\mathbf{z}\in\mathbb{R}^{8}$. The latent vector is then encoded into the quantum state for subsequent variational evolution.

This formulation provides a controlled transition from the classical feature space to the 8 qubits Hilbert space. It also reduces the number of classical degrees of freedom before quantum processing.

\begin{equation}
\mathbf{z} = \frac{\pi}{2}\tanh(\mathbf{W}\mathbf{x} + \mathbf{b})
\end{equation}

In standard neural networks, the activation function mainly introduces nonlinearity. In the proposed framework, we assign a different physical interpretation to the mapped vector $\mathbf{z}$. We treat its components as canonical coordinates in the quantum phase space. These coordinates prepare the classical features for quantum encoding and subsequent rotations on the Bloch sphere.

We also interpret the dimensionality reduction layer as a discrete step in the renormalization group (RG) flow. This mapping reduces the classical feature space by removing degrees of freedom that are not relevant to the learning task. It retains the relevant latent variables represented by the eight parameters. These variables determine the macroscopic outputs of the system, including the semantic class and geometric coordinates.

The control parameters are constrained to the interval $[-\frac{\pi}{2},\frac{\pi}{2}]$ before quantum encoding. This range avoids the redundant representations introduced by the $2\pi$ periodicity of quantum rotation gates. It also reduces ambiguity in the encoded quantum states.

We use a hyperbolic tangent function to project the classical features into the defined angular range. This transformation preserves the distinguishability of the classical feature vectors during quantum encoding. It therefore maintains their representation within the Hilbert space. This approach follows the geometric principles discussed in \cite{ref52} for quantum representation learning.

The quantum circuit is initialized in the state $|\psi_0\rangle = |0\rangle^{\otimes N}$. Hadamard gates are then applied to all qubits. Each qubit is transformed into the state $|+\rangle = \frac{1}{\sqrt{2}}(|0\rangle + |1\rangle)$, which places the state on the equator of the Bloch sphere. The probabilities of measuring $|0\rangle$ and $|1\rangle$ are therefore equal.

This initialization represents a state of maximum uncertainty before the network receives any informative features. When the input feature vector is zero, the circuit has no basis for favoring a particular class or spatial coordinate. The initial state is therefore unbiased with respect to the classification and localization tasks. This provides a symmetric starting point for gradient based optimization, allowing small changes in the input features to evolve the quantum state toward representations specific to each task.

After the Hadamard gates are applied to all qubits, the compressed classical features are encoded into the Hilbert space using a unitary angle encoding scheme based on $R_Y$ rotation gates. The input vector $\mathbf{z}$ therefore determines the resulting parameterized quantum state. In this state, the classical information is represented through the amplitudes and relative phases of the 256 computational basis states.

\begin{equation}
|\psi_{in}(\mathbf{z})\rangle = R_Y(\mathbf{z})H|0\rangle^{\otimes N} = \bigotimes_{k=0}^{N-1} e^{-i\frac{\mathbf{z}_k}{2}\sigma_y^{(k)}} H^{(k)} |0\rangle_k
\end{equation}

Here, $\sigma_y^{(k)}$ denotes the Pauli Y operator acting on the $k$th qubit.

\subsection{Quantum Circuit (Ansatz)}
The proposed quantum circuit follows the hardware efficient ansatz (HEA) design strategy described in \cite{ref53}. The unitary evolution $U(\theta)$ is constructed from $L=6$ layers. Each layer consists of an entangling operation followed by single qubit $R_Y$ and $R_Z$ rotations.

\begin{equation}
U(\theta) = \prod_{l=1}^{L=6} U_l(\theta_l) = \prod_{l=1}^{L=6} \left[ \left( \bigotimes_{k=0}^{N-1} R_Z(\theta_{l,k}^{Z})R_Y(\theta_{l,k}^{Y}) \right) \mathcal{U}_{ent} \right]
\end{equation}

The rotation operators are represented by $\bigotimes_{k=0}^{N-1} e^{-i\frac{\theta_{l,k}}{2}\hat{\sigma}}$, where $\hat{\sigma} \in \{Y,Z\}$ denotes the corresponding Pauli operator. The tensor product represents the parallel application of the parameterized single qubit rotations to each qubit. The parameters $\theta_{l,k}$ serve as trainable weights of the network and allow the model to explore the Hilbert space through single qubit rotations.

To generate multipartite entanglement, we use nearest neighbor entangling operations. We consider two circuit topologies: linear and circular. The entangling unitary is defined as

\begin{equation}
\mathcal{U}_{ent} = \prod_{(m,n)\in E} CNOT_{m,n} = \prod_{(m,n)\in E} e^{-i\frac{\pi}{4}(I-\sigma_z^{(m)})(I-\sigma_x^{(n)})}
\end{equation}

Here, $E$ denotes the set of connected qubit pairs. The indices $m$ and $n$ identify the control and target qubits of each CNOT gate. In the linear topology, the entangling operations connect adjacent qubits. In the circular topology, the last qubit is also connected to the first qubit. This additional connection closes the interaction graph and imposes periodic boundary conditions.

The central computational component of our model is a variational quantum ansatz. Its variational structure is analogous to the ansatz used in the variational quantum eigensolver (VQE) framework. We interpret the $L=6$ layer circuit as a physical system undergoing coherent evolution under a time dependent Hamiltonian $H(\theta,t)$. Each layer $l$ applies a specific unitary evolution $U_l(\theta_l)$. This unitary transformation evolves the initial wavefunction through the Hilbert space.

Within this framework, the trainability of the model is analyzed through its energy landscape. This landscape is characterized by the geometry of the loss function $\mathcal{L}(\theta)$. By optimizing the parameters $\theta$, the network learns a specific loss trajectory that maps the input feature state to a target state. The relevant information for classification and localization is concentrated in the low energy region associated with the global minimum of the loss landscape. The Pauli-Z observables provide the measurement basis for the final quantum predictions.

The evolution of the system wavefunction $|\psi(\theta)\rangle$ represents a transition from informational uncertainty to structural order. We interpret the $U(\theta)$ layers as successive stages of the RG flow. At each stage, irrelevant information is progressively eliminated, while relevant information is preserved and amplified. This information corresponds to the identity and spatial location of the object.

During this evolution, the entanglement structure changes and the amplitude distribution becomes concentrated on basis states relevant to the task. This process concentrates the wavefunction representation on information relevant to the task. The final measurement in the Pauli Z basis therefore provides information about the object identity and spatial coordinates. The circuit consequently extracts structured information from the high-dimensional input manifold.

\subsection{Algebra of Observables and Projective Measurements}
The readout mechanism is defined by the composite observable  $\widehat{O} = \bigoplus_k \alpha_k \sigma_z^{(k)}$, which represents a direct sum of local measurement operators. From a physical perspective, this structure corresponds to independent quantum sensors acting on individual qubits. The readout therefore extracts information from local measurements on individual qubits.

Here $\sigma_z^{(k)}$ denotes the Pauli Z operator acting on the $k$th qubit. It projects the final wavefunction onto the computational basis. The coefficients $\alpha_k \in \mathbb{R}$ serve as classical scaling parameters. The expectation value of the observable $\sigma_z^{(k)}$ is obtained from

\begin{equation}
\langle \sigma_z^{(k)} \rangle = Tr\left( \sigma_z^{(k)} \rho_{out} \right) = \langle \psi_{in}(\mathbf{z}) | U^{\dagger}(\theta)\sigma_z^{(k)}U(\theta) | \psi_{in}(\mathbf{z}) \rangle
\end{equation}

Here, $\rho_{out}$ denotes the output density matrix after the variational quantum evolution. The expectation values obtained from the local Z measurements form the basis of the final readout.

The design of the readout mechanism uses a composite observable operator and follows an approach termed observable learning. The quantum circuit performs local measurements of the Pauli Z observables $\sigma_z^{(k)}$. The classical neural network then learns the optimal coefficients $\alpha_k$ that combine these measurements into an effective observable operator.

This learned operator is used to discriminate between semantic classes, such as dog and cat, and to predict the bounding box coordinates. This approach reduces the computational burden on the quantum circuit by shifting part of the readout optimization to the classical network. It also helps prevent the emergence of barren plateaus during circuit training.

\subsection{Task Specific Partitioning of the Hilbert Space}
The Hilbert space is structured as the tensor product of two task-specific Hilbert spaces: $\mathcal{H} = \mathcal{H}_{class} \bigotimes \mathcal{H}_{bbox}$. The two subspaces represent different aspects of the object detection task. The classification subspace captures semantic information. The localization subspace captures geometric information. This Partitioning separates the representation associated with classification from the geometric representation associated with localization.

The classification subspace $\mathcal{H}_{class}$, defined by the first four qubits $k \in \{0,1,2,3\}$, encodes the semantic information required for object classification. The variational quantum circuit transforms the input feature vector into a quantum state. Its local expectation values and inter qubit correlations contain information relevant to the target class. The Pauli Z expectation values obtained from the classification qubits are mapped to class logits through the learned classical readout layer. A Softmax function then converts these logits into normalized class probabilities.

The localization subspace $H_{bbox}$, defined by the four qubits $k \in \{4,5,6,7\}$, serves as a quantum regressor. This formulation enables continuous regression through the continuous dependence of quantum expectation values on the circuit parameters. Specifically, the state of each qubit is continuously parameterized on the Bloch sphere, governed by the unitary evolution $U(\theta)$. While a single projective measurement yields discrete binary outcomes $\{-1, +1\}$, the corresponding expectation value  $\langle \sigma_z^{(k)} \rangle$ is a continuously differentiable function of the input parameters, bounded in $[-1, 1]$.
To bridge this quantum continuous representation of Hybrid Quantum Regression with macroscopic visual features, we map these observables to the spatial coordinate space $\{x, y, w, h\}$ using an affine transformation. Crucially, while the coefficients of this mapping are optimized dynamically via observable learning, the mapping itself remains affine. This transformation provides a linear coordinate mapping from the quantum observables to the bounding box coordinates, extracting bounding box coordinates without introducing any non-linear classical learning overhead, thereby ensuring that the core regression logic is executed within the quantum state space

\subsection{Loss Function}
In this paper, the optimization is formulated as a Multi Task Learning (MTL) problem. This framework requires a balance between two distinct objectives. The first objective is translation invariance for classification. The second objective is translation equivariance for localization. The network therefore minimizes two loss functions simultaneously. The categorical cross entropy loss $\mathcal{L}_{CE}$ is used to optimize the discrete classification task. The regression loss $\mathcal{L}_{GIoU}$, based on the Generalized Intersection over Union (GIoU), is used for localization. The total loss is defined as:

\begin{equation}
\mathcal{L}_{total} = \mathcal{L}_{CE}(c,\hat{c}) + \lambda \mathcal{L}_{GIoU}(b,\hat{b}) \label{eq:total}
\end{equation}

Here, $\hat{c}$ and $\hat{b}$ denote the target class and the ground truth bounding box, respectively. The variables $c$ and $b$ represent the corresponding outputs of the proposed model. We set $\lambda=10$ as a gradient scaling hyperparameter that regulates the Pareto optimization trajectory is there an reference here?. This weighting term balances the contributions of the two tasks during training. The classification gradients tend to converge more rapidly and can therefore dominate the optimization dynamics. The weighting term therefore ensures that the localization task receives sufficient gradient updates during backpropagation. This balance supports the co evolution of semantic and geometric features within the shared Hilbert space.

The classification objective is defined by the standard cross entropy loss:

\begin{equation}
\mathcal{L}_{CE} = - \sum_i \hat{c}_i \log(c_i)
\end{equation}

For spatial localization, the geometric loss is defined using the Generalized Intersection over Union (GIoU):

\begin{equation}
\mathcal{L}_{GIoU} = 1 - IoU + \frac{|C / U|}{|C|}
\end{equation}

Here, $C$ denotes the minimum enclosing convex region containing both bounding boxes, and $U$ denotes their union. The Intersection over Union IoU provides the basic measure of spatial overlap. It is defined as the intersection area of the predicted and ground truth bounding boxes divided by their union:

\begin{equation}
IoU = \frac{| b \cap \hat{b} |}{| b \cup \hat{b} |}
\end{equation}

\subsection{Classical Optimizer}
The optimization procedure in the proposed architecture uses an end to end backpropagation pipeline. The gradient of the total loss is propagated from the quantum observables through the trainable classical-quantum interface. The architecture is end to end differentiable. This property allows the gradient to propagate across the classical and quantum components of the model.

We adopt the quantum transfer learning paradigm by freezing the internal weights of the classical backbone. This preserves the visual features learned during pretraining. The architecture remains differentiable across the classical-quantum interface. The gradients propagate through the quantum observables and update the compression mapping layer. The classical projection of the features is therefore optimized during training to improve compatibility with the quantum Hilbert space embedding. This process is enabled by the differentiable connection between the classical encoder, the quantum processor, and the classical decoder.

The parameter shift rule \cite{ref54} enables the analytical evaluation of quantum gradients within the classical computational graph. These gradients are then propagated through the quantum transfer learning model using the chain rule:

\begin{equation}
\frac{\partial\mathcal{L}}{\partial W_{CNN}} = \sum_{k=0}^{N-1} \frac{\partial\mathcal{L}}{\partial f} \cdot \frac{\partial f}{\partial \langle Z_k \rangle} \cdot \frac{\partial \langle Z_k \rangle}{\partial \theta_Q} \cdot \frac{\partial \theta_Q}{\partial W_{CNN}}
\end{equation}

The term $\frac{\partial\mathcal{L}}{\partial f} \cdot \frac{\partial f}{\partial \langle Z_k \rangle}$ corresponds to the output section of the model. It describes the propagation of the loss gradient through the classical readout layer and quantifies how changes in the quantum expectation values affect the loss function. The final output layer maps the expectation values of qubits 0 to 3 to the semantic classes and the expectation values of qubits 4 to 7 to the bounding box coordinates. The resulting outputs are compared with the corresponding ground truth values through the loss function.

The term $\frac{\partial \langle Z_k \rangle}{\partial \theta_Q}$ describes the sensitivity of the quantum observables to the trainable parameters of the variational quantum circuit. The parameter-shift rule forms an integral part of the backpropagation pipeline. The quantum gradients are evaluated using the parameter-shift rule. For the $R_Y$ and $R_Z$ rotation gates used in the circuit, the corresponding derivatives can be evaluated at shifted parameter values. For the standard generators of these gates, the required shifts are $\theta_Q \pm \pi/2$, yielding the gradient as the difference between the two shifted evaluations multiplied by $1/2$.

This analytical procedure integrates the quantum processor into the backpropagation computational graph without relying on finite difference approximations.

The term $\frac{\partial \theta_Q}{\partial W_{CNN}}$ pertains to the trainable interfacial layer that transforms the 1280 dimensional MobileNet features into the rotation angles of the variational quantum circuit. This layer was previously designated as the compression mapping.

\subsection{Quantum Metrics}
In this paper, we define four fundamental metrics to monitor the dynamics of the variational quantum circuit throughout the training phase. These metrics provide a quantitative description of the quantum state during optimization.

\subsubsection{Total von Neumann Entropy}
We calculate the total von Neumann entropy from the density matrix of the complete 8 qubit system, denoted by $\rho_{total}$. It is defined as

\begin{equation}
S(\rho_{total}) = - Tr(\rho_{total} \log_2 \rho_{total}) = - \sum_i \lambda_i \log_2 \lambda_i
\end{equation}
where $\lambda_i$ denote the eigenvalues of $\rho_{total}$.

Our simulations use a state vector formalism. The complete system therefore remains in a pure state throughout the simulation. From the perspective of quantum mechanics, the von Neumann entropy of a pure state is zero: $S(\rho_{total}) = 0$. If the same system is implemented on physical Noisy Intermediate Scale Quantum (NISQ) hardware, this metric can serve as a quantitative indicator of the loss of state purity caused by hardware noise and decoherence. In that setting, a nonzero entropy can indicate a transition from a pure state to a mixed state.

\subsubsection{Bipartite von Neumann Entropy}
This metric quantifies the bipartite entanglement between the classification subsystem, defined by qubits 0 to 3, and the localization subsystem, defined by qubits 4 to 7. An increase in this entropy indicates stronger quantum informational coupling between the classification and localization subsystems. This result indicates that the two task specific subsystems are entangled.

The global quantum system remains in a pure state. Therefore $Tr(\rho_{total}^{2}) = 1$ and $S(\rho_{total}) = 0$.

For a bipartite pure state, the von Neumann entropies of the two complementary subsystems are identical: $S(\rho_{cls}) = S(\rho_{bbox})$. We therefore quantify the bipartite entropy by computing $S(\rho_{cls})$.

The reduced density matrix of the classification subsystem, $\rho_{cls}$ is obtained by taking the partial trace over the localization subsystem:

\begin{equation}
\rho_{cls} = Tr_{bbox}(\rho_{total})
\end{equation}

The von Neumann entropy of the classification subsystem is then calculated from this reduced density matrix:

\begin{equation}
S(\rho_{cls}) = - Tr(\rho_{cls} \log_2 \rho_{cls})
\end{equation}

The theoretical upper bound of the von Neumann entropy is $S \in [0, N_{sub}]$, where $N_{sub}$ denotes the number of qubits in the considered subsystem. For our four qubit partition, the entropy is bounded by $S \in [0,4]$ in units of bits because the logarithm is defined with base 2.

An entropy of $S = 0$ corresponds to a pure reduced state and indicates zero bipartite entanglement between the classification and localization subsystems. Conversely, $S = 4$ corresponds to a maximally mixed reduced state and represents maximal bipartite entanglement for the four qubit subsystem.

\subsubsection{Meyer-Wallach Entanglement Measure}
This metric quantifies multipartite entanglement by measuring the average entanglement of each qubit with the remainder of the system. It therefore characterizes the extent to which individual qubits are entangled with the rest of the system. For an N qubit system, with $N=8$ in our case, the Meyer- Wallach measure $Q$ is defined as

\begin{equation}
Q(|\psi\rangle) = \frac{2}{N}\sum_{k=0}^{N-1} (1 - Tr(\rho_k^2))
\end{equation}

where $\rho_k$ denotes the reduced density matrix of the $k$th qubit. It is obtained by tracing out the remaining qubits:

\begin{equation}
\rho_k = 
Tr_{i \neq k}(|\psi\rangle\langle\psi|)
\end{equation}

The theoretical range of the Meyer-Wallach measure is $Q \in [0,1]$, and $Q$ is dimensionless. A value of $Q=0$ corresponds to a completely separable state with no multipartite entanglement. A value of $Q=1$ corresponds to maximal multipartite entanglement.

\subsubsection{Total Shannon Entropy}
Before measurement, the 8 qubit system occupies a pure entangled state in a Hilbert space of dimension $2^8 = 256$. Within this unitary regime, the von Neumann entropy of the global system is zero: $S(\rho_{total}) = 0$.

Upon measurement in the computational basis using the Pauli Z operator, the wavefunction collapses. At this stage, the quantum state gives rise to a classical probability distribution over the computational basis outcomes. The Shannon entropy $H$ is then calculated using Eq.~\eqref{eq:shannon}:

\begin{equation}
H = - \sum_x P(x)\log_2 P(x)
\label{eq:shannon}
\end{equation}

The total Shannon entropy quantifies the uncertainty of the classical probability distribution generated by measurement in the computational basis.

\subsection{Thermodynamic Interpretation and Learning Phase Transition}
In the object detection task, the proposed hybrid neural network predicts the joint probability distribution of the true class $c$ and the true bounding box coordinates $b$, conditioned on the input image and the extracted feature vector $\mathbf{x}$. Based on the Maximum Likelihood Estimation (MLE) principle, the probability of convergence to the ground truth is expressed through the exponential of the negative loss functions \cite{ref55}.

For classification, the loss function is the cross entropy loss $\mathcal{L}_{CE}$:

\begin{equation}
P(c | \mathbf{x}) = \exp(-\mathcal{L}_{CE})
\end{equation}

In the classification subspace, the affine linear layer produces a set of logits $z_i$. These logits are passed through a Softmax activation function to obtain the probability of each class, dog or cat:

\begin{equation}
P_i = \frac{\exp(z_i)}{\sum_j \exp(z_j)}
\label{eq:softmax}
\end{equation}

In statistical mechanics, the Boltzmann distribution gives the probability of finding a system in a microstate with energy $E_i$ at temperature $T$:

\begin{equation}
P_i = \frac{1}{Z_{thermo}}\exp\left( -\frac{E_i}{k_B T} \right)
\label{eq:boltzmann}
\end{equation}

where $Z_{thermo}$ denotes the partition function. By comparing Eqs.~\eqref{eq:softmax} and \eqref{eq:boltzmann}, we obtain

\begin{equation}
E_i = - z_i \cdot k_B T
\end{equation}

The logits can therefore be interpreted as effective energy levels $E_i$ in the classification subsystem. Because the logit $z_i$ is dimensionless, whereas the energy $E_i$ has units of energy, information theoretic natural units are adopted to ensure dimensional consistency between the Softmax and Boltzmann distributions. Within this convention, the product of the Boltzmann constant and the effective temperature is normalized such that $k_B T = 1$. This normalization is maintained throughout the subsequent derivation.

For localization, the loss function is $\mathcal{L}_{GIoU}$. The geometric probability is formulated as a Boltzmann distribution:

\begin{equation}
P(b | \mathbf{x}) = \frac{1}{Z_{bbox}}\exp\left( -\lambda \mathcal{L}_{GIoU} \right)
\label{eq:bboxprob}
\end{equation}
Assuming conditional independence between the geometric and classification losses as a computational assumption, the joint probability of the predicted system state is expressed according to Eq. (\ref{eq:total}) as
\begin{align}
	P\left(c,b\middle|\mathbf{x}\right)
	&=
	\frac{1}{Z_{\mathrm{bbox}}}
	\exp\left[
	-\left(
	\mathcal{L}_{\mathrm{CE}}
	+
	\lambda\mathcal{L}_{\mathrm{GIoU}}
	\right)
	\right] \nonumber \\
	&=
	\frac{1}{Z_{\mathrm{bbox}}}
	\exp\left(
	-\mathcal{L}_{\mathrm{total}}
	\right).
	\label{eq:join prob}
\end{align}
where $Z_{\mathrm{bbox}}$ is the normalization constant in the continuous localization space, and $\lambda=\frac{1}{T_{\mathrm{eff}}}$ is the weighting coefficient. Equation (\ref{eq:boltzmann}) describes the system at the microscopic scale. At the macroscopic scale, represented by the integrated output of the neural network, the probability of a state can be related to its effective Helmholtz free energy.
\begin{equation}
	P_{\mathrm{Macro}}
	=
	\frac{1}{Z_{\mathrm{thermo}}}
	\exp\left(
	-\frac{F}{k_{\mathrm{B}}T}
	\right).
	\label{eq:macro prob}
\end{equation}
where $P_{\mathrm{Macro}}$ denotes the probability of the macroscopic state, and F is the effective Helmholtz free energy. In the proposed hybrid architecture, the integrated network output represents the joint prediction of class c and bounding box coordinates b. This output is treated as the macroscopic state of the system within the thermodynamic interpretation. Therefore,
\begin{equation}
	P_{Macro} =
	P\left(c,b\middle|\mathbf{x}\right).
	\label{eq:marco state}
\end{equation}	
The reference predictive state corresponds to the ground truth in the learning space. It is interpreted as the ground state at the global minimum of the effective free energy landscape. By defining the reference state and dividing Eq. (\ref{eq:marco state}) by the probability of the reference state, we obtain
\begin{equation}
	\frac{P_{Macro}}{P_\mathrm{ref}}=\frac{P\left(c,b\middle|\mathbf{x}\right)}{{P}_\mathrm{ref}}
\end{equation}	
Using the probability formulation in Eq. (\ref{eq:join prob}) and the thermodynamic relation in Eq. (\ref{eq:macro prob}), the two descriptions can be compared through the corresponding probability ratio:
\begin{equation}
	\exp{\left(-\frac{\Delta F}{k_{B} T}\right)}=\exp{\left(-\Delta {\mathcal{L}}\right)}
\end{equation}	
So, we obtain
\begin{equation}
	\frac{\Delta F}{k_{B} T} =\Delta \mathcal{L}=\mathcal{L}_{total}-\mathcal{L}_\mathrm{ref}
\end{equation}
Given $\mathcal{L}_{\mathrm{ref}}=0$ and $k_B T=1$, we obtain
\begin{equation}
	\Delta F=\mathcal{L}_{total}
\end{equation}
Therefore, in the proposed architecture, the total learning loss can be interpreted as the excess Helmholtz free energy relative to the reference state. During training, the hybrid neural network evolves through the parametric space toward a stable state. For a system with fixed volume, a fixed number of qubits, and a fixed number of trainable parameters, Helmholtz free energy provides the appropriate thermodynamic potential for this interpretation.

Based on the classical Helmholtz relation for isothermal systems, $F = U - TS$, the total loss is decomposed into two physical components with distinct roles.

The first component is the effective internal energy $U \equiv \lambda \mathcal{L}_{GIoU}$. The geometric localization loss represents the effective internal energy and structural potential of the system. Minimizing this term through backpropagation is interpreted as performing virtual mechanical work that reduces the geometric potential associated with the spatial mismatch between the predicted and ground truth bounding boxes. This process promotes greater overlap between the two boxes while reducing the uncovered region within the smallest enclosing geometric region. Consequently, the predicted bounding boxes move toward spatial alignment with the ground truth.

The second component is the effective entropic term $-TS \equiv \mathcal{L}_{CE}$. This term is interpreted as an effective entropic contribution within the proposed thermodynamic analogy. The semantic loss, represented by cross entropy, describes the magnitude of informational disorder and uncertainty. By minimizing the classification loss and the associated Kullback-Leibler (KL) divergence, the network moves toward a more ordered predictive state.

From the perspective of Information Geometry, minimizing the cross entropy in the discrete classification subspace, defined by qubits 0 to 3, is directly related to reducing the KL divergence $D_{KL}(q \| P)$.

In machine learning theory, the network output distribution $P(c | \mathbf{x})$ is optimized against the empirical data distribution or ground truth distribution $q$ by minimizing the KL divergence. The cross entropy loss is mathematically decomposed as

\begin{equation}
\mathcal{L}_{CE} = H(q) + D_{KL}(q \| P)
\end{equation}

Because the classification labels in the dataset are deterministic one hot vectors, the entropy of the target distribution is zero: $H(q) = 0$.

Therefore, the cross entropy loss becomes

\begin{equation}
\mathcal{L}_{CE} = D_{KL}(q \| P)
\end{equation}

To connect this information theoretic quantity with statistical mechanics, the predicted class distribution is written in Boltzmann form $P_i = \frac{\exp(z_i)}{\sum_j \exp(z_j)} = \frac{e^{-E_i}}{Z}$. In the thermodynamic formalism, the partition function is related to the Helmholtz free energy by $F = -\ln Z$. Since the target distribution $q_i$ is equal to one only for the correct class $i = \hat{c}$ and zero for all other classes, the KL divergence becomes

\begin{equation}
D_{KL}(q \| P) = \sum_i q_i \ln \frac{q_i}{P_i} = \ln(1) - \ln P_{i=\hat{c}} = E_{\hat{c}} + \ln Z
\end{equation}

By substituting the Helmholtz free energy relation, we obtain

\begin{equation}
D_{KL}(q \| P) = E_{\hat{c}} - F
\end{equation}

The KL divergence therefore serves as a measure of the excess free energy. Its value depends on the energy associated with the correct class state and the free energy of the system. This relation shows how the energy of the correct class is positioned relative to the total free energy.

In contrast, the continuous localization subsystem, defined by qubits 4 to 7, represents a continuous configuration space. The outputs of this subsystem are interpreted as the most probable state, or mode, of an implicit Boltzmann probability distribution in the geometric state space. The probability of finding the target bounding box at a point in this configuration space is described by Eq.~\eqref{eq:bboxprob}.

Because the regression space is continuous, the normalization constant is expressed as an integral rather than a discrete sum. The corresponding partition function is

\begin{equation}
Z_{bbox} = \int_{\mathbb{R}^4} \exp\left( -\frac{\mathcal{L}_{GIoU}(b,\hat{b})}{T_{eff}} \right) db
\end{equation}

Applying a logarithmic transformation to the continuous partition function $Z_{bbox}$ recovers the Helmholtz free energy relation within the localization space. The optimization of the regression space can therefore be interpreted as a reduction in Helmholtz free energy toward thermodynamic equilibrium.

The learning process in the proposed 8 qubit variational circuit therefore follows a coupled thermodynamic trajectory. In the adopted information theoretic natural units, this trajectory is described by the unified phase transition relation.

\begin{equation}
\Delta F_{total} = \Delta U_{geom} - T\Delta S_{semantic} \equiv \lambda \mathcal{L}_{GIoU} + \mathcal{L}_{CE}
\end{equation}

During optimization, the Adam optimizer is interpreted as an anisotropic adaptive thermal bath. The effective temperature $T$ associated with each parameter in the quantum circuit is interpreted as the corresponding learning rate. This parameter is adjusted independently and adaptively during the optimization process.

Within this thermodynamic interpretation, the Adam optimizer acts as a thermodynamic engine. By reducing the informational entropy associated with the semantic subsystem and the effective internal energy of the spatial subsystem, it evolves the quantum system from its initial state toward an effective equilibrium state within the energy landscape.

\section{Results and Discussion} \label{sec4}
The proposed model was implemented in Python. The model was executed on a laptop equipped with a 13th generation Intel Core i5 13420H processor and 16 GB of RAM operating at 5200 MT/s. The model was evaluated using two entanglement topologies, circular and linear. As shown in Table~\ref{tab:topology}, the model with the circular CNOT topology achieved the best empirical results.
\begin{widetext}
\begin{center}
\begin{table}[H]
 			\centering
			\caption{Topological dependence of quantum information metrics and detection performance.}
			\label{tab:topology}
			\renewcommand{\arraystretch}{1.25}
			\setlength{\tabcolsep}{7pt}
			\begin{tabular}{lccccc}
				\toprule \hline \hline
				\textbf{CNOT Topology of the Proposed Model}
				& \textbf{Bipartite von Neumann Entropy}
				& \textbf{Meyer--Wallach Measure ($Q$)}
				& \textbf{mAP@0.5}
				& \textbf{mAP@0.5:0.95} \\
				\midrule
				\hline \hline
				 \hspace{20mm} Circular
				& 0.1688
				& 0.0370
				& 0.9448
				& 0.6558 \\
				
				\hspace{20mm} Linear
				& 0.7067
				& 0.3622
				& 0.9264
				& 0.6453 \\
				
				\bottomrule
			\end{tabular}
	\end{table}
\end{center}
\end{widetext}

The empirical data in Table~\ref{tab:topology} provide evidence for the quantum information bottleneck phenomenon in the NISQ era. This analysis extends beyond conventional performance benchmarking of different architectures. Within this theoretical framework, the quantum circuit is treated as a physical system. The flow of information between the classification and localization subsystems is controlled by CNOT operators.

Considering the theoretical upper bound of $S_{max} = \log_2(2^4) = 4$ bits for the reduced entropy in this 8 qubit system, under a symmetric partition into two 4 qubit subsystems, the results in this table reveal a nontrivial phenomenon in information physics. Achieving high object detection performance in computer vision does not require the full capacity of the Hilbert space or maximal entanglement. Instead, it requires topological and targeted constrained entanglement. Instead, it requires topological and purposefully constrained entanglement.

The thermodynamic and informational behavior of the system can be analyzed across two distinct structural regimes.

\textbf{Circular Topology (Periodic Boundary Conditions):}By imposing periodic boundary conditions, this structure introduces a symmetric connectivity pattern into the system. Within this architecture, the circuit acts as an information bottleneck and ﬁlter. It suppresses information that is less relevant to the learning tasks and allows the subsystems to evolve in a nearly pure state. The architecture exhibits a low bipartite von Neumann entropy of 0.1688 bits and a constrained Meyer-Wallach entanglement measure of 0.0370. From a topological perspective, this low-entanglement configuration is associated with a low-energy region of the optimization landscape. The circular topology achieves strong object detection performance, with an mAP@0.5:0.95 of 0.6558.

\textbf{Linear Topology (Open Boundary Conditions):} In contrast, the linear circuit lacks boundary symmetry. This structure produces a higher degree of global entanglement and a higher bipartite von Neumann entropy of 0.7067 bits. From the perspective of quantum information theory, this increase in entropy indicates stronger entanglement between the classification and localization subsystems. The linear circuit also acts as an information bottleneck and achieves competitive performance. 

However, comparative analysis shows that the circular circuit provides stronger feature filtering and task disentanglement. It also achieves superior object detection performance.

\begin{figure}[H]\centering\includegraphics[width=0.9\linewidth]{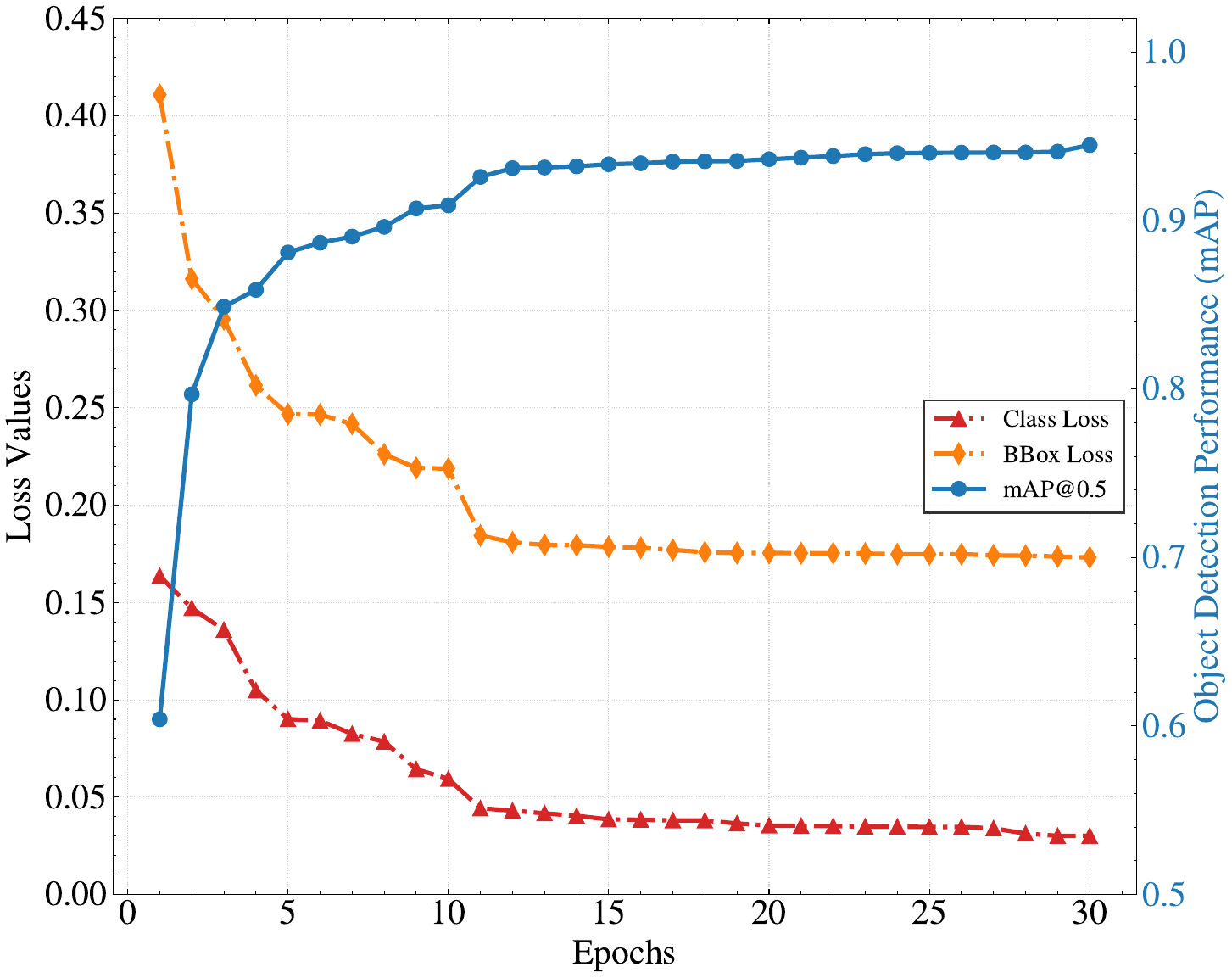}\caption{Optimization dynamics and Pareto equilibrium of the hybrid multi-task framework with the circular CNOT topology.}\label{fig:optimization}\end{figure}

Examining the model behavior with circular entanglement in Fig.~\ref{fig:optimization} reveals a close connection between statistical mechanics and quantum machine learning. In this architecture, the 8 qubit variational quantum circuit uses 96 rotational parameters as training weights. The circuit evolves the quantum state across a hypersurface in the geometric space of the loss function. The optimization objective is to guide the system toward the ground state. This state corresponds to the minimum energy level and represents the optimal solution to the object detection problem.

In classical multi task learning approaches, the simultaneous optimization of regression and classification can lead to negative transfer. In this setting, the gradients of one task can conflict with those of the other. However, the symmetry and synchronized decay of both loss functions in Fig.~\ref{fig:optimization} indicate coordinated optimization by the Dressed Quantum Circuit (DQC) head.

This performance advantage is not simply the result of learning the two tasks independently. It can instead be interpreted as a geometric phase transition within the Hilbert space. The entangled circuit structure establishes a shared representation space in which information renormalization reduces conflicts between the two tasks. This interaction allows the tasks to benefit from shared information rather than compete during optimization.

From a thermodynamic perspective, the values on the left vertical axis, representing the loss values, are not interpreted solely as conventional engineering errors. In the proposed framework, the total loss function is interpreted as the Helmholtz free energy in information theoretic natural units. The optimization algorithm therefore acts as an anisotropic and adaptive thermal bath that evolves the system toward the minimum of this free energy. The decay of the loss functions can be interpreted through two physical components.

The bounding box loss curve, denoted as BBox Loss and shown by orange diamonds, represents the effective internal energy of the system. The geometric loss function is interpreted as a potential well that defines the geometric boundaries of the object. The marked decrease of this quantity from approximately 0.41 to 0.18 indicates that the continuous distribution moves toward the global minimum of this potential well.

The classification loss curve, denoted as Class Loss and shown by red triangles, represents the effective entropic contribution of the system. In this interpretation, the output logits define the corresponding energy levels. The rapid convergence of this quantity toward 0.03 indicates the crystallization of structural information and the minimization of uncertainty.

\begin{figure}[H]\centering\includegraphics[width=0.9\linewidth]{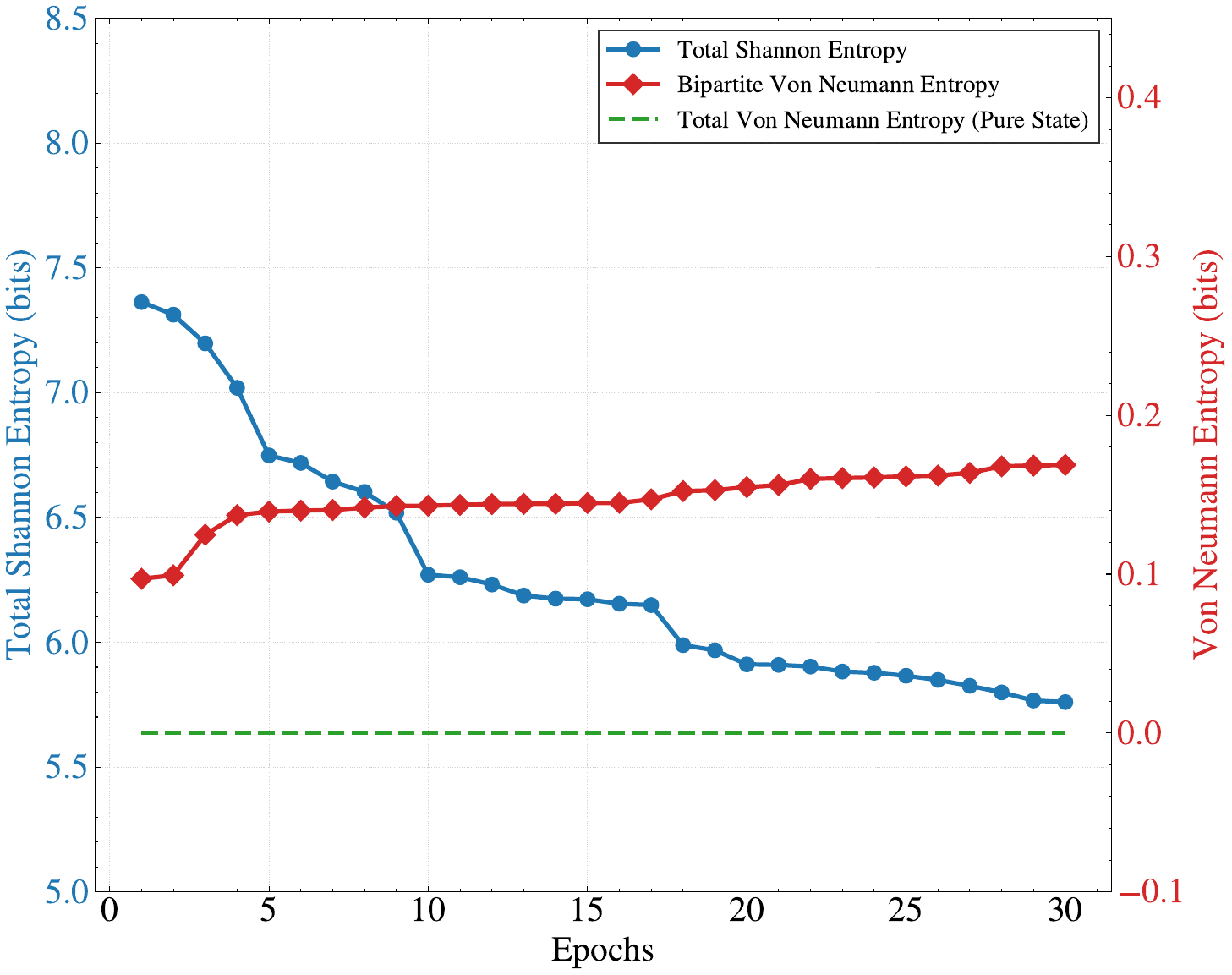}\caption{Thermodynamic cooling and entanglement generation during optimization with the circular CNOT topology.}\label{fig:cooling}\end{figure}

In evaluating the proposed model with the circular entanglement structure, the physical behavior of the system during training is examined at two levels. The first level describes macroscopic classical transitions. The second level describes microscopic quantum dynamics. Figure~\ref{fig:cooling} reveals three fundamental phenomena in the evolution of this hybrid architecture:

\textbf{I. Thermodynamic Cooling (Blue Curve).} The total Shannon entropy curve describes the macroscopic behavior of the system in the presence of a thermal bath. This quantity decreases continuously from approximately 7.4 bits during the initial epochs to 5.8 bits at the end of training. This decrease represents the thermodynamic cooling behavior of the system within the proposed framework. The optimization process is interpreted as the minimization of the Helmholtz free energy in information theoretic natural units. In this interpretation, the optimization algorithm acts as an anisotropic and adaptive thermal bath rather than only as a numerical optimizer. The continuous reduction in Shannon entropy indicates a decrease in shot noise and statistical uncertainty. The stochastic information state of the system therefore evolves toward a more deterministic and structured information state.

\textbf{II. Inter-Subsystem Correlations (Red Curve).} The most distinct non classical phenomenon in this architecture is the increase in the bipartite von Neumann entropy. The reduced density matrix of the classification subsystem shows that the entanglement between the classification subsystem, defined by qubits 0 to 3, and the geometric localization subsystem, defined by qubits 4 to 7, begins near 0.1 bits. It undergoes a rapid phase transition during the first 5 epochs. The entropy then stabilizes near 0.17 bits. This physical behavior contrasts with the heuristic assumption of conditional independence between tasks used in traditional networks. Within the proposed thermodynamic interpretation, the classification loss and the spatial potential well become coupled through the quantum circuit. The geometric  coordinates of the object and its semantic identity are therefore correlated across the Hilbert space. This interaction forms a unified feature representation for the two tasks.

\textbf{III. Unitarity and Quantum Isolation (Green Dashed Line).} The total von Neumann entropy remains at 0.0 throughout the 30 epochs of training. This behavior is consistent with the unitary evolution of the circuit. The system therefore remains in a pure state throughout the learning process: $S(\rho_{total}) = 0$. The unitarity condition is also preserved: $Tr(\rho^2) = 1$. This result indicates that the information flow and gradients within this part of the architecture are preserved under the noise free simulation conditions. Consequently, no increase in entropy attributable to environmental decoherence or numerical instabilities is observed during backpropagation.

\begin{figure}[H]\centering\includegraphics[width=0.9\linewidth]{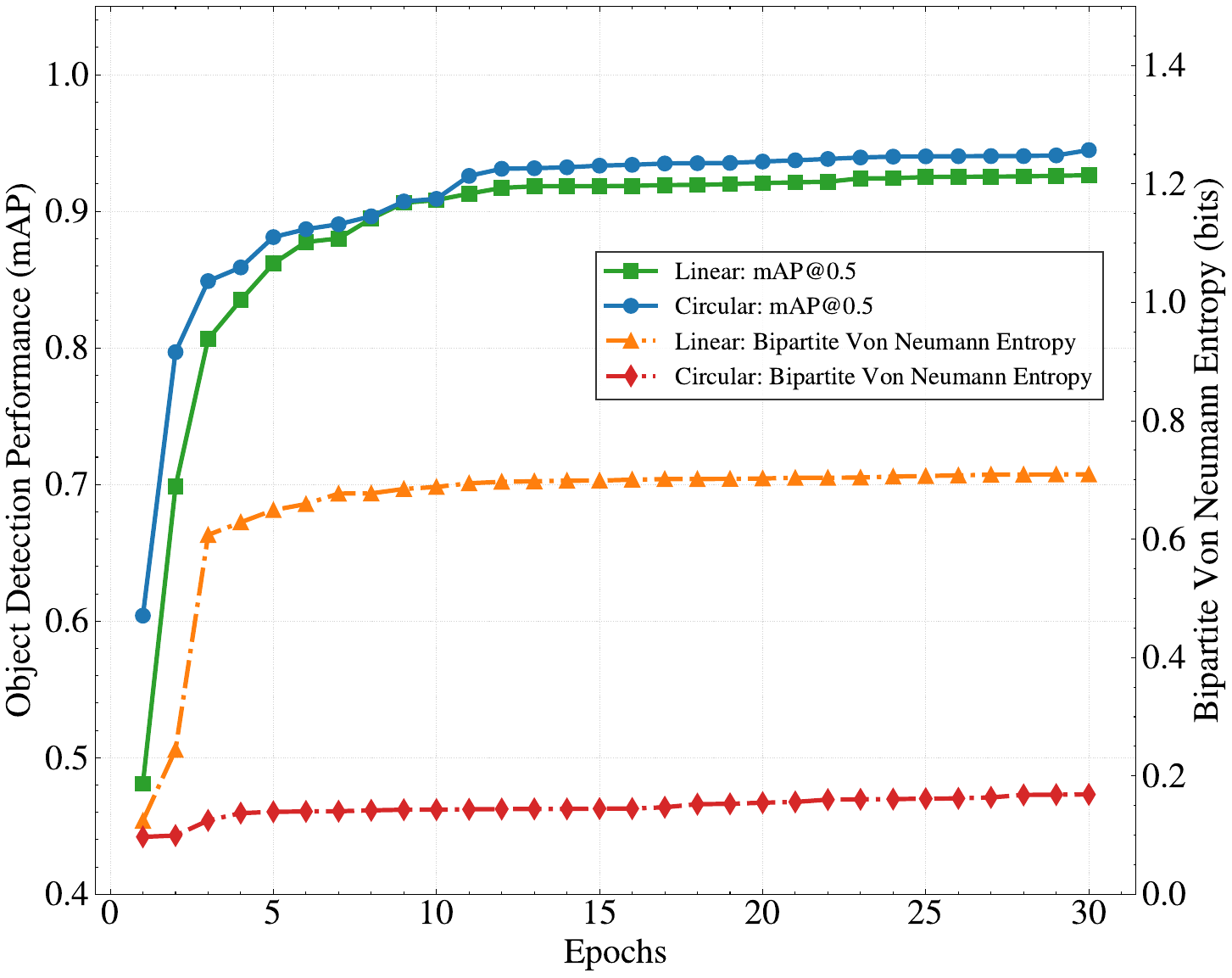}\caption{Topological advantage of circular entanglement architectures in hybrid object detection.}\label{fig:topology}\end{figure}

The analysis of the system behavior in Fig.~\ref{fig:topology} reveals an important  shift in the design of quantum neural networks. For the 8 qubit system partitioned into two 4 qubit subsystems, the theoretical upper bound of the reduced entropy is 4 bits. The computed values for the linear and circular architectures are 0.7 and 0.16 bits, respectively. Both values remain substantially below this theoretical bound.

However, both architectures achieve highly competitive object detection performance. This performance is evaluated using mAP and is comparable with that of advanced classical models.

These observations show that, contrary to conventional expectations, solving complex real world problems such as object detection does not require full use of the available state space or the generation of maximally entangled states. Instead, the network performance is associated with a targeted restriction of entanglement.

Circuits that approach maximal entanglement can approach an approximate unitary 2-design and may exhibit vanishing gradients. In contrast, maintaining the entropy at low values, corresponding to 17.5\% of the available capacity in the linear model and only 4\% in the circular model, indicates that the learning process operates within a structured physical subspace. This restricted subspace preserves the optimization dynamics while maintaining competitive object detection performance.

The most striking phenomenon in this circuit is the counterintuitive behavior of the bipartite von Neumann entropy during the transition from the linear to the circular topology. The circular topology introduces an additional direct connection between the two subsystems through a CNOT gate, with qubit 7 as the control and qubit 0 as the target. Such an additional connection might be expected to increase the entanglement between the subsystems. However, the bipartite von Neumann entropy decreases from 0.7 to 0.16 bits.

This behavior is therefore associated with the specific topological structure of the circular circuit. The underlying mechanisms can be analyzed through three physical and topological effects:

\textbf{I.} In the linear topology, the absence of structural connectivity at the two ends of the chain, namely qubits 0 and 7, produces pronounced edge effects. As the Adam optimizer minimizes the loss function, it must operate under these connectivity constraints. The open boundaries limit information transfer between the two ends of the array. To compensate for this limitation and facilitate information flow across the chain, the network tends toward a more globally entangled configuration. Closing the loop to form the circular topology removes this structural constraint. The system can then perform the object detection task without requiring additional global entanglement to compensate for boundary effects.

\textbf{II.} In the linear topology, information transfer between the two subsystems is mediated through a single physical bottleneck at the boundary between qubits 3 and 4. The additional connection introduced by the circular topology closes the loop and provides a second pathway for information propagation. By optimizing the rotational parameters, the network can exploit destructive phase interference. This interference can suppress entanglement generated along one pathway through the phase structure of the other. As a result, irrelevant information is attenuated, effectively filtering task irrelevant components from the learned representation.

\textbf{III.} The two tasks have distinct objectives. The first concerns the semantic identity of the object, or ``what''. The second concerns its spatial localization, or ``where''. These objectives are optimized using the Cross Entropy and GIoU loss functions, respectively. This separation promotes structural decoupling between the corresponding subsystems.

Because of its limited structural connectivity, the linear model appears to remain near a low-energy region associated with a bipartite entanglement entropy of 0.7 bits. In contrast, the circular topology introduces additional degrees of freedom within the free energy landscape. These degrees of freedom provide an optimization pathway enabled by the circuit topology. This pathway guides the system toward a lower energy optimum. This pathway guides the system toward a lower energy optimum. At this point, the learned feature representations become substantially decoupled. The Meyer-Wallach entanglement measure also decreases markedly to 0.03.

\begin{figure}[H]\centering\includegraphics[width=0.9\linewidth]{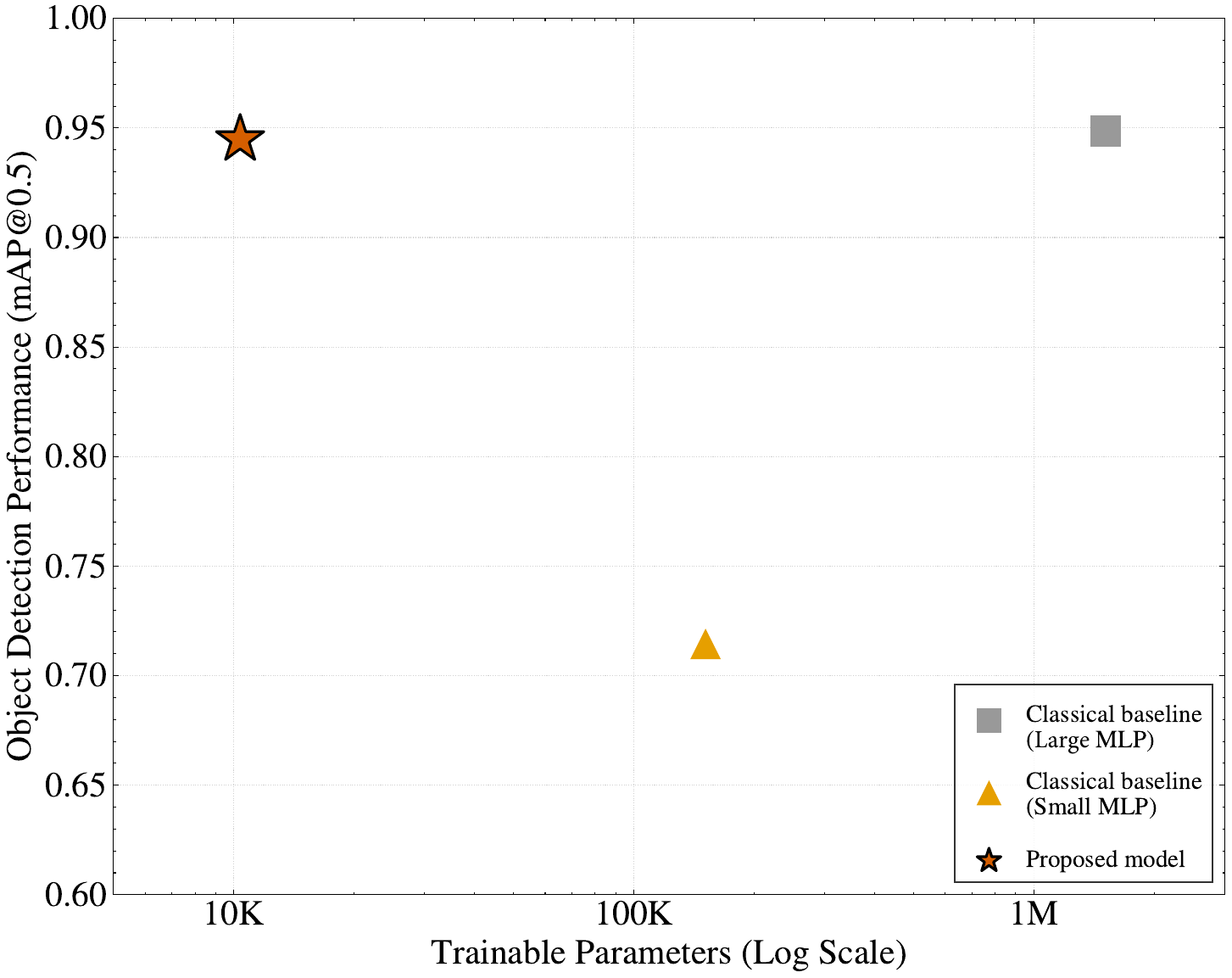}\caption{The parameter efficiency frontier of hybrid classical-quantum architectures.}\label{fig:efficiency}
\end{figure}

\begin{widetext}
\begin{center}

\begin{table}[H]
	\centering
	\caption{Parameter efficiency and performance of the entanglement-enhanced architecture.}
	\label{tab:params}
	\renewcommand{\arraystretch}{1.25}
	\setlength{\tabcolsep}{6pt}
	\begin{tabular}{lrrrr}
		\toprule \hline \hline
		\textbf{Model}
		& \textbf{Trainable Parameters}
		& \textbf{Parameter Reduction (\%)}
		& \textbf{mAP@0.5}
		& \textbf{mAP@0.5:0.95} \\
		\midrule \hline \hline
		
		Classical baseline (Large MLP)
		& 1,509,638
		& --
		& 0.9486
		& 0.6561 \\
		
		Classical baseline (Small MLP)
		& 150,964
		& 90.00
		& 0.7145
		& 0.5111 \\
		
		\textbf{Proposed Model (Circular Entanglement Structure)}
		& \textbf{10,366}
		& \textbf{99.31}
		& \textbf{0.9448}
		& \textbf{0.6558} \\
		
		\bottomrule
	\end{tabular}
\end{table}
\end{center}
\end{widetext}
To provide a fair comparison of parameter efficiency, the pre-trained MobileNetV2 backbone is kept identical and fully frozen across all evaluated models. Its parameters are therefore treated as a fixed computational component and are excluded from the trainable parameter counts reported in Table \ref{tab:params}.
The reported parameter countsand the corresponding $99.31 \%$ reduction  refer exclusively to the task specific prediction heads. These heads consist of the classical large MLP architectures and the Dressed Quantum Circuit (DQC) head. This controlled comparison provides a direct comparison of parameter efﬁciency between the classical and hybrid classical-quantum architectures. It also highlights the difference in their representation spaces. The classical head operates through trainable parameters in a Euclidean parameter space, whereas the variational quantum circuit evolves quantum states in Hilbert space.

Figure~\ref{fig:efficiency} provides empirical evidence for the physical realization of the theoretical potential expected from NISQ era architectures. The analysis extends beyond a direct performance comparison. The presented results shift the efficiency frontier and indicate a paradigm shift from classical overparameterization toward quantum expressivity.

In the deep learning literature, traditional neural networks rely on overparameterization to disentangle nonlinear correlations between competing tasks, such as semantic classification and geometric localization. The classical baseline model, Large MLP, uses more than 1.5 million parameters to achieve a mAP@0.5 of 0.9486. Reducing the parameter count by 90\% results in the Small MLP model with 150,964 parameters. This model exhibits a capacity limitation, with its mAP@0.5 decreasing sharply to 0.7145. This pronounced performance degradation indicates that standard classical architectures lack the inductive bias required to preserve spatial and semantic features effectively in low parameter regimes.

The proposed model achieves a 99.31\% reduction in the number of parameters while using only 10,366 trainable parameters. It maintains near equivalent localization performance. The stringent mAP@0.5:0.95 metric evaluates localization quality across increasingly strict IoU thresholds. The large classical model achieves an mAP@0.5:0.95 of 0.6561, whereas the hybrid model achieves 0.6558. This close agreement represents a substantial shift toward the upper left region of the efficiency frontier in Fig.~\ref{fig:efficiency}. These results indicate that the hybrid model preserves the learned features despite the substantial reduction in parameter count. Instead, these features are represented within a high-dimensional quantum state space.

The 99.31\% reduction in trainable parameters observed in this study is not the result of algebraic compression, such as classical pruning techniques. Instead, it arises from implementing the learned representation within a tensor product Hilbert space. Classical models represent feature transformations through computationally intensive matrix operations. In contrast, the proposed model exploits a circular entanglement structure.

In this architecture, the variational gates do not primarily function as generators of new features. Instead, they act as phase control operators and steer the quantum state through an exponentially large state space. This structure enables high expressivity with a minimal parameter footprint.

The near equivalent performance of a model with approximately ten thousand trainable parameters and a model with 1.5 million trainable parameters across complex computer vision metrics can be interpreted as a manifestation of information renormalization. Within this structure, entanglement acts as an informational mechanism that couples task relevant representations. The semantic identity of the object and its geometric coordinates are encoded within the quantum bottleneck, while irrelevant information is progressively filtered. 

\begin{figure}[H]
	\begin{center}
	\includegraphics[width=1\linewidth]{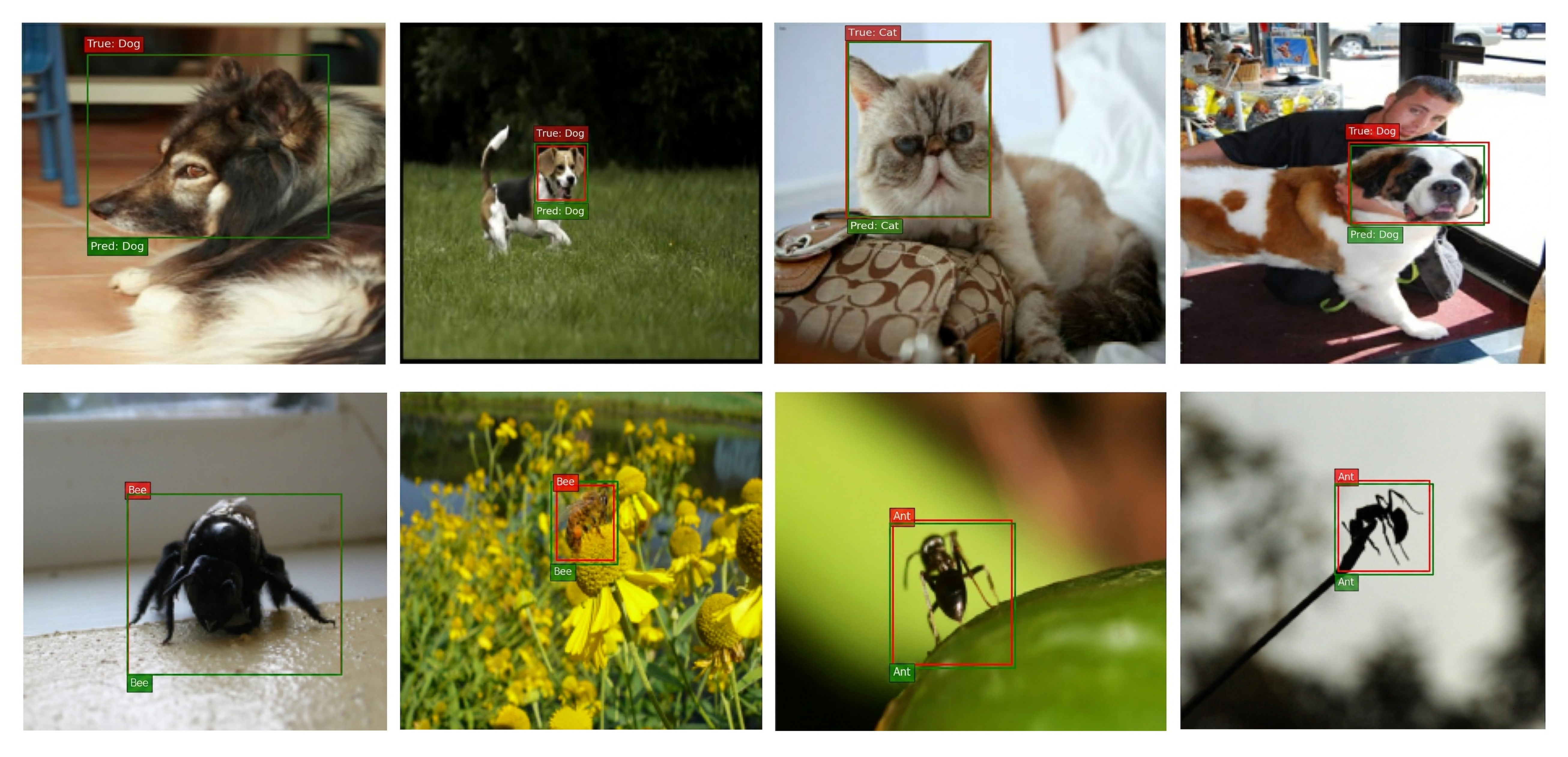}\caption{Detection results visualization using proposed hybrid classical-quantum object detection model (the green box stands for ``Prediction'', the red box shows ``True Label'').}\label{fig:detection}
	\end{center}
\end{figure}

The outputs of the proposed model, evaluated on the Kaggle dataset \cite{ref51}, are visualized in Fig.~\ref{fig:detection}. The model demonstrates high object detection performance. The visual difference between the ground truth labels and the model predictions is negligible. All results were computed and reported using the Kaggle dataset. The ant and bee dataset was adopted from the first quantum transfer learning study \cite{ref5}.

\section{Conclusion and Future Works} \label{sec5}
In this paper, we presented a hybrid classical-quantum transfer learning architecture for object detection based on an 8 qubit variational quantum circuit with linear and circular entanglement topologies. The proposed framework extends quantum transfer learning beyond conventional classification. It jointly learns semantic classification and continuous geometric localization within a task aware Hilbert space. The Hilbert space is partitioned into two task-specific quantum subspaces. Local quantum expectation values are then used for the final readout. This design enables hybrid prediction of bounding box coordinates.

The results show that circuit topology provides an effective mechanism for regulating information flow and entanglement. High detection performance does not require maximal use of the available Hilbert space or maximal entanglement. The circular topology maintains competitive detection performance while substantially reducing both entanglement and the number of trainable parameters.

Based on the methodology and results presented in this study, the following insights are obtained:

\begin{itemize}
\item \textbf{Thermodynamic interpretation of learning:} The learning process can be interpreted through changes in Helmholtz free energy, with GIoU and cross entropy contributing to effective internal energy and entropic terms, respectively. The decrease in Shannon entropy indicates thermodynamic cooling, as the model evolves from a state of higher information uncertainty toward a more structured representation.
\item \textbf{Information renormalization:} The quantum circuit can be interpreted as an information renormalization process that progressively transforms high-dimensional features into task relevant representations. Quantum embedding, entanglement dynamics, and quantum interference collectively suppress task irrelevant information while preserving task relevant features.
\item \textbf{Quantum information bottleneck:} High performance in complex computer vision tasks does not require full use of the Hilbert space or maximal entanglement. Instead, relevant information can be selectively filtered through controlled and limited entanglement.
\item \textbf{Parameter efficiency:} The proposed hybrid architecture achieves competitive object detection performance with only 10,366 trainable parameters, corresponding to a 99.31\% reduction relative to the classical baseline.
\end{itemize}

Together, these findings indicate that the key resource in a multi task quantum model is not the magnitude of entanglement alone. Rather, it is the ability to organize and regulate quantum correlations according to the structure of the learning task. These results support a design principle based on task aware organization of the Hilbert space and topology controlled information flow. Targeted entanglement can therefore be more effective than entanglement maximization for learning task relevant representations.

\subsection{Future Works}
In future studies, this hybrid architecture can be extended to multimodal learning for open vocabulary object detection. A parameterized quantum circuit can replace the classical text encoder. This extension could examine whether quantum embeddings can align visual features from MobileNetV2 with semantic text descriptions in a shared Hilbert space. Future work can also explore Quantum Generative Adversarial Networks (QGANs) and Quantum Diffusion Models to improve robustness under occlusion and varying lighting conditions. These models can generate high fidelity quantum entangled feature representations for rare object classes. Such representations can provide a physically informed data augmentation strategy for the detection model. Additionally, future work could investigate the theoretical dynamics of gradients during the topological phase transition. Particular attention can be given to the evolution of the Fubini-Study metric during training. This analysis can provide a deeper understanding of the relationship between circuit topology, training dynamics, and quantum information flow.


\end{document}